\documentclass[english,conference]{IEEEtran}
\usepackage[T1]{fontenc}
\usepackage[utf8]{inputenc}
\usepackage{float}
\usepackage{mathrsfs}
\usepackage{mathtools}
\usepackage{amsmath}
\usepackage{amsthm}
\usepackage{amssymb}
\usepackage{graphicx}

\makeatletter

\providecommand{\tabularnewline}{\\}
\floatstyle{ruled}
\newfloat{algorithm}{tbp}{loa}
\providecommand{\algorithmname}{Algorithm}
\floatname{algorithm}{\protect\algorithmname}

\theoremstyle{plain}
\newtheorem{thm}{\protect\theoremname}
\theoremstyle{plain}
\newtheorem{lem}[thm]{\protect\lemmaname}

\usepackage{eqnarray}
\usepackage{mathrsfs}
\usepackage{epsfig}
\usepackage[english]{babel}
\usepackage{subfigure}
\usepackage{epstopdf}
\usepackage{import}
\usepackage{color}
\usepackage{colortbl}\usepackage{cite}
\usepackage{algorithm}

\usepackage{bbm}
\usepackage{cases}
\usepackage{array}
\usepackage[english]{babel}
\usepackage{times}
\usepackage{color}
\usepackage{amsfonts}
\usepackage{psfrag}
\usepackage{fancyhdr}
 \usepackage{algorithmic}
\allowdisplaybreaks

\usepackage{dsfont}
\usepackage{babel}
\providecommand{\lemmaname}{Lemma}
\providecommand{\theoremname}{Theorem}
\makeatother

\usepackage{babel}
\providecommand{\lemmaname}{Lemma}
\providecommand{\theoremname}{Theorem}

\begin{document}

\title{Ultra-Reliable Low-Latency Vehicular Networks: Taming the Age of
Information Tail}

\author{\IEEEauthorblockN{Mohamed~K.~Abdel-Aziz\IEEEauthorrefmark{1}, Chen-Feng~Liu\IEEEauthorrefmark{1},
Sumudu~Samarakoon\IEEEauthorrefmark{1}, Mehdi~Bennis\IEEEauthorrefmark{1},
and Walid~Saad\IEEEauthorrefmark{2} }\IEEEauthorblockA{\IEEEauthorrefmark{1}Centre for Wireless Communications, University
of Oulu, Finland \\
\IEEEauthorrefmark{2}Wireless@VT, Bradley Department of Electrical
and Computer Engineering, Virginia Tech, Blacksburg, VA, USA \\
E-mails: \{mohamed.abdelaziz, chen-feng.liu, sumudu.samarakoon, mehdi.bennis\}@oulu.fi,
walids@vt.edu }}
\maketitle
\begin{abstract}
While the notion of \emph{age of information} (AoI) has recently emerged
as an important concept for analyzing ultra-reliable low-latency communications
(URLLC), the majority of the existing works have focused on the \emph{average}
AoI measure. However, an average AoI based design falls short in properly
characterizing the performance of URLLC systems as it cannot account
for extreme events that occur with very low probabilities. In contrast,
in this paper, the main objective is to go beyond the traditional
notion of average AoI by characterizing and optimizing a URLLC system
while capturing the AoI tail distribution. In particular, the problem
of vehicles' power minimization while ensuring stringent latency and
reliability constraints in terms of probabilistic AoI is studied.
To this end, a novel and efficient mapping between both AoI and queue
length distributions is proposed. Subsequently, extreme value theory
(EVT) and Lyapunov optimization techniques are adopted to formulate
and solve the problem. Simulation results shows a nearly two-fold
improvement in terms of shortening the tail of the AoI distribution
compared to a baseline whose design is based on the maximum queue
length among vehicles, when the number of vehicular user equipment
(VUE) pairs is $80$. The results also show that this performance
gain increases significantly as the number of VUE pairs increases. 
\end{abstract}


\section{Introduction\label{sec:Introduction}}

Vehicle-to-vehicle (V2V) communication will play an important role
in the next generation (5G) mobile networks, and is envisioned as
one of the most promising enabler for intelligent transportation systems
\cite{Araniti2013,Zeng2018,8255748}. Typically, V2V safety applications
(forward collision warning, blind spot/lane change warning, and adaptive
cruise control) are known to be \emph{time-critical}, as they rely
on acquiring real-time status updates from individual vehicles. In
this regard, the European telecommunications standards institute (ETSI)
has standardized two safety messages: cooperative awareness message
(CAM) and decentralized environmental notification message (DENM)
\cite{etsi2011intelligent}. One key challenge in V2V networks is
to deliver ultra-reliable and low-latency communications for such
status update messages.

Indeed, achieving ultra-reliable low-latency communication (URLLC)
represents one of the major challenges facing 5G and vehicular networks
\cite{Mehdi_riskTail}. In particular, a system design based on conventional
average values is not adequate to capture the URLLC requirements,
since averaging often ignores the occurrence of extreme events (e.g.,
high latency events) that negatively impact the overall performance.
To overcome this challenge, one can resort to the robust framework
of \emph{extreme value theory} (EVT) that can allow a full characterization
of the probability distributions of extreme events, defined as the
tail of the latency distribution or queue length \cite{EVT}. Remarkably,
the majority of the existing V2V literature which address latency
and reliability, focus only on average performance metrics \cite{Ashraf2016,Ikram2017,Liu2017,Mei2018,Cristina},
which is not sufficient to enable URLLC. Only a handful of recent
works have considered extreme values for vehicular networks \cite{Mouradian2016,ChenBaseline2018}.
In particular, the work in \cite{Mouradian2016} focuses on studying
large delays in vehicular networks using EVT, via simulations using
realistic mobility traces, without considering any analytical formulations.
In \cite{ChenBaseline2018}, the authors study the problem of transmit
power minimization subject to a new reliability measure in terms of
maximal queue length among all vehicle pairs. Therein, EVT was utilized
to characterize the maximal queue length.

Since V2V safety applications are time-critical, the freshness of
a vehicle's status updates is of high importance \cite{champatiAoI,AoI_First}.
A relevant metric in quantifying this freshness is the notion of \emph{age
of information} (AoI) proposed in \cite{AoI_First}. AoI is defined
as the time elapsed since the generation of the latest status update
received at a destination. Thus, providing quality-of-service (QoS)
guarantees in terms of AoI is essential for any time-critical application.
It should be noted that, minimizing AoI is fundamentally different
from delay minimization or throughput maximization. In \cite{AoI_First},
the authors derive the minimum AoI at an optimal operating point that
lies between the extremes of maximum throughput and minimum delay.
Recently, minimizing the \emph{average} AoI in vehicular networks
was studied (e.g., see \cite{AoI_First,AoIBaiocchi2017,Zhou2018,AoIAbd-Elmagid2018},
and references therein). However, while interesting, a system design
based on average AoI cannot enable the unique requirements of URLLC.
Instead, the AoI distribution needs to be considered especially when
dealing with time-critical V2V safety applications.

The main contribution of this paper is to go beyond the conventional
average AoI notion by developing a novel framework to characterize
and optimize the tail of AoI in vehicular networks. In particular,
our key goal is to enable vehicular user equipment (VUE) pairs to
minimize their transmit power while ensuring stringent latency and
reliability constraints based on a probabilistic AoI measure. To this
end, we first derive a novel relationship between the probabilistic
AoI and the queue length of each VUE. Then, we use the fundamental
concepts of EVT to characterize the tail and excess value of the vehicles'
queues, which is then incorporated as a constraint within our optimization
problem. To solve the formulated problem, a roadside unit (RSU) is
used to cluster VUEs into disjoint groups, thus mitigating interference
and reducing the signaling overhead between VUEs and RSU. Subsequently,
by leveraging Lyapunov stochastic optimization techniques, each VUE
can locally optimize its power subject to probabilistic AoI constraints.
Simulation results corroborate the usefulness of EVT in characterizing
the distribution of AoI. The results also show over two-fold performance
gains in AoI and queue length compared to a baseline whose design
is based on the maximum queue length \cite{ChenBaseline2018}. Our
results also expose an interesting tradeoff between the arrival rate
of the status updates and the average and worst AoI achieved by the
network.

The rest of this paper is organized as follows. In Section \ref{sec:System-model},
the system model is presented. The reliability constraints and the
studied problem are formulated in Section \ref{sec:Problem-formulation},
followed by the proposed AoI-aware resource allocation policy in Section
\ref{sec:AoI-aware-optimal-control}. In Section \ref{sec:Numerical-results},
numerical results are presented followed by conclusions in Section
\ref{sec:Conclusion}.

\section{System Model}

\label{sec:System-model} 
\begin{figure}[t]
\centering \includegraphics[width=1\columnwidth]{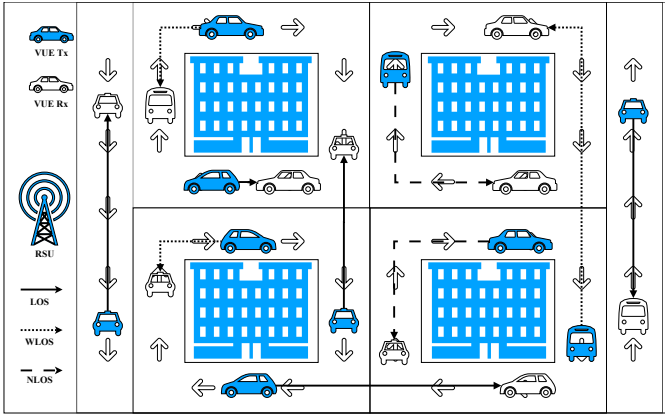}
\caption{System and path loss models of the considered V2V communication network.}
\label{fig:Mobility_and_Path_model} 
\end{figure}

As shown in Fig.~\ref{fig:Mobility_and_Path_model}, we consider
a V2V communication that uses a Manhattan mobility model \cite{Manhattan},
which is composed of a set $\mathcal{K}$ of $K$ VUE transmitter-receiver
pairs under the coverage of a single RSU. During the entire communication
lifetime, the association of each transmitter-receiver is fixed. We
consider a slotted communication timeline which is indexed by $t$,
and the duration of each slot is denoted by $\tau$. Additionally,
all VUE pairs share a set $\mathcal{N}$ of $N$ orthogonal resource
blocks (RBs) with bandwidth $\omega$ per RB. We further denote the
RB usage as $\eta_{k}^{n}(t)\in\left\{ 0,1\right\} ,\forall k\in\mathcal{K},n\in\mathcal{N}$,
in which $\eta_{k}^{n}(t)=1$ indicates that RB $n$ is used by VUE
pair $k$ in time slot $t$. Otherwise, $\eta_{k}^{n}(t)=0$. The
transmitter of pair $k$ allocates a transmit power $P_{k}^{n}(t)\geq0$
over RB $n$ to serve its receiver subject to $\sum_{n\in\mathcal{N}}\eta_{k}^{n}(t)P_{k}^{n}(t)\leq P_{\textrm{max}}$,
where $P_{\textrm{max}}$ is the total power budget. Moreover, $h_{kk'}^{n}(t)$
is the instantaneous channel gain, including path loss and channel
fading, from the transmitter of pair $k$ to the receiver of pair
$k'$ over RB $n$ in slot $t$. We consider the $5.9\text{ GHz}$
carrier frequency and adopt the path loss model in \cite{Path_loss_model}.

For our model, we express an arbitrary transmitter's and an arbitrary
receiver's Euclidean coordinates as $\boldsymbol{x}=(x_{i},x_{j})\in\mathbb{R}^{2}$
and $\boldsymbol{y}=(y_{i},y_{j})\in\mathbb{R}^{2}$, respectively.
When the transmitter and receiver are on the same lane, we consider
a line-of-sight (LOS) path loss value $l_{0}\left\Vert \boldsymbol{x}-\boldsymbol{y}\right\Vert ^{-\alpha}$,
where $\left\Vert .\right\Vert $ is the $l_{2}$-norm, $l_{0}$ is
the path loss coefficient, and $\alpha$ is the path loss exponent.
If the transmitter and receiver are located separately on perpendicular
lanes with one near the intersection within a distance $\mathscr{D}$,
then we consider a weak-line-of-sight (WLOS) path loss model $l_{0}\left\Vert \boldsymbol{x}-\boldsymbol{y}\right\Vert _{1}^{-\alpha}$
with the $l_{1}$-norm $\left\Vert .\right\Vert _{1}$. Finally, if
both transmitter and receiver are located on the perpendicular lanes,
but the distances to the intersection are larger than $\mathscr{D}$,
then we adopt a non-line-of-sight (NLOS) path loss value $l_{0}^{'}\left(|x_{i}-y_{i}|.|x_{j}-y_{j}|\right)^{-\alpha}$,
with the path loss coefficient $l_{0}^{'}<l_{0}\left(\frac{\mathscr{D}}{2}\right)^{\alpha}$.
Fig.~\ref{fig:Mobility_and_Path_model} briefly illustrates these
three path loss cases. For this network, the data rate of VUE pair
$k$ in time slot $t$ (in the unit of packets per slot) is expressed
as 
\begin{equation}
R_{k}(t)=\frac{\tau}{Z}{\textstyle \sum\limits _{n\in\mathcal{N}}}\omega\log_{2}\left(1+\frac{P_{k}^{n}(t)h_{kk}^{n}(t)}{N_{0}\omega+I_{k}^{n}(t)}\right),\label{eq:Rate}
\end{equation}
with $Z$ the packet length in bits and $N_{0}$ is the power spectral
density of the additive white Gaussian noise. Here, $I_{k}^{n}(t)=\sum_{k'\in\mathcal{K}/k}\eta_{k'}^{n}(t)P_{k'}^{n}(t)h_{k'k}^{n}(t)$
indicates the received aggregate interference at the receiver of VUE
pair $k$ over RB $n$ from other VUE pairs operating over the same
RB $n$.

Furthermore, each VUE transmitter has a queue buffer to store the
data to be delivered to the desired receiver. Denoting VUE pair $k$'s
queue length at the beginning of slot $t$ as $Q_{k}(t)$, the queue
dynamics are given by, 
\begin{equation}
Q_{k}(t+1)=\max\left(Q_{k}(t)-R_{k}(t),0\right)+A,\label{eq:Physical_queue}
\end{equation}
where $A$ is the constant packet arrival rate per slot, under the
assumption of deterministic periodic arrivals. Hence, packet $i$'s
arrival time instance can be denoted as $\frac{i}{A}\tau$. It can
be noted that the indices of the packets that arrive during slot $t$
satisfy $\ensuremath{i\in[tA,(t+1)A-1]}$, while the packets that
are served during the same slot fulfill 
\begin{equation}
tA-Q_{k}(t)\leq i\leq tA-1-\max\left(Q_{k}(t)-R_{k}(t),0\right).\label{eq:served_indices}
\end{equation}

\section{Reliability Constraints based on Age of Information\label{sec:Problem-formulation}}

Providing real-time status updates in mission critical applications
(e.g., CAMs) is a key use case for V2V networks. Further, these applications
rely on the ``freshness'' of the data, which can be quantified by
the concept of AoI \cite{AoI_First}: 
\begin{equation}
\Delta_{k}(T)\triangleq T-\max_{i}\left(T_{k}^{\text{A}}\left(i\right)\mid T_{k}^{\text{D}}\left(i\right)\leq T\right).
\end{equation}
Here, $\Delta_{k}(T)$ is the AoI of VUE pair $k$ at a time instant
$T$. $T_{k}^{\text{A}}\left(i\right)$ and $T_{k}^{\text{D}}\left(i\right)$
denotes the arrival and departure instant of packet $i$ of VUE pair
$k$, respectively. As a reliability requirement, we impose a probabilistic
constraint on the AoI for each VUE pair $k\in\mathcal{K}$, i.e.,
\begin{equation}
\lim_{T\rightarrow\infty}\textrm{Pr}\left\{ \Delta_{k}(T)>d\right\} \leq\epsilon_{k}\text{, }\forall k\in\mathcal{K},\label{eq:AoI_constraint}
\end{equation}
where $d$ is the age limit, and $\epsilon_{k}\ll1$ is the tolerable
AoI violation probability. It was shown in \cite{champatiAoI} that
for a given age limit $d$ with $\frac{A}{\tau}\geq\frac{1}{d},$
the steady state distribution of AoI for a D/G/1 queue can be characterized
as, 
\begin{equation}
\lim_{T\rightarrow\infty}\textrm{Pr}\Bigl\{\Delta_{k}(T)>d\Bigr\}=\lim_{T\rightarrow\infty}\textrm{Pr}\left\{ T_{k}^{\text{D}}(\hat{i})>T\right\} ,\label{eq:ZubaidyTheorem}
\end{equation}
where $\hat{i}\triangleq\lceil\frac{A}{\tau}(T-d)\rceil$ is the index
of the packet that first arrives at or just after time $T-d$. Next,
in Lemma \ref{lem:Assuming-that-packet } we propose a mapping between
the steady state distribution of the departure instant of a given
packet and the queue length. 
\begin{lem}
\label{lem:Assuming-that-packet }Assuming that $T$ is observed at
the beginning of each slot $t+1$, i.e., $T=\tau(t+1)$, then 
\[
\textrm{Pr}\left\{ T_{k}^{\text{D}}(\hat{i})>T\right\} \leq\textrm{Pr}\left\{ Q_{k}(t)>R_{k}(t)-\psi\right\} ,
\]
will be satisfied with $\psi=2-(\frac{d}{\tau}-1)A$. 
\end{lem}
\begin{IEEEproof}
Since $T=\tau(t+1)$, then $\textrm{Pr}\left\{ T_{k}^{\text{D}}(\hat{i})>T\right\} =\textrm{Pr}\left\{ \hat{i}\text{ is NOT served before time }\tau(t+1)\right\} $,
which means that $\hat{i}$ is not served at or before time slot $t$.
Subsequently, we apply \eqref{eq:served_indices} and derive 
\begin{align*}
 & \textrm{Pr}\left\{ T_{k}^{\text{D}}(\hat{i})>T\right\} \\
 & \overset{(a)}{=}\textrm{Pr}\left\{ \hat{i}>tA-1-\max\left(Q_{k}(t)-R_{k}(t),0\right)\right\} \\
 & \leq\textrm{Pr}\left\{ \frac{A}{\tau}(\tau(t+1)-d)+1>tA-1-\left(Q_{k}(t)-R_{k}(t)\right)\right\} \\
 & =\textrm{Pr}\left\{ Q_{k}(t)>R_{k}(t)-\psi\right\} ,
\end{align*}
where $\hat{i}=\lceil\frac{A}{\tau}(T-d)\rceil\leq\frac{A}{\tau}(\tau(t+1)-d)+1$
is used in step \emph{(a)}. It also should be noted that if $\hat{i}$
departs after $\tau(t+1)$, then $Q_{k}(t)-R_{k}(t)>0$, which is
used in the same step. 
\end{IEEEproof}
Combining the results of Lemma \ref{lem:Assuming-that-packet } and
\eqref{eq:ZubaidyTheorem}, the probabilistic constraint \eqref{eq:AoI_constraint}
can be rewritten as, 
\begin{equation}
\lim_{C\to\infty}\frac{1}{C}{\textstyle \sum\limits _{t=0}^{C-1}}\textrm{Pr}\left\{ Q_{k}(t)>R_{k}(t)-\psi\right\} \leq\epsilon_{k}\text{, }\forall k\in\mathcal{K}.\label{eq:equiv_queue_constraint}
\end{equation}
As previously discussed, enabling URLLC requires the characterization
of the tail of the AoI distribution. Therefore, we further investigate
the event $Q_{k}(t)>R_{k}(t)-\psi$ and study the tail behavior of
AoI in the following part. First, we need to introduce the \emph{Pickands–Balkema–de
Haan theorem} for exceedances over threshold \cite{EVT}. Consider
a random variable $Q$ whose cumulative distribution function (CDF)
is denoted by $F_{Q}(q)$. As a threshold $\delta$ closely approaches
$F_{Q}^{-1}(1)$, the conditional CDF of the excess value $X=Q-\delta>0$
is $F_{X\text{|}Q>\delta}(x)\approx G\left(x;\sigma,\xi\right)$ where
\[
G\left(x;\sigma,\xi\right)=\begin{cases}
1-(\max\{1+\frac{\xi x}{\sigma},0\})^{-\frac{1}{\xi}}, & \xi\neq0,\\
1-e^{-\frac{x}{\sigma}}, & \xi=0.
\end{cases}
\]
Here, $G\left(x;\sigma,\xi\right)$ is the generalized Pareto distribution
(GPD) whose mean and variance are $\frac{\sigma}{1-\xi}$ and $\frac{\sigma^{2}}{(1-\xi)^{2}(1-2\xi)}$,
respectively. Moreover, the characteristics of the GPD depend on the
scale parameter $\sigma>0$ and the shape parameter $\xi<\frac{1}{2}$.

\setcounter{equation}{13} 
\begin{figure*}
\begin{flalign}
J_{k}^{(X)}(t+1) & =\max\left(J_{k}^{(X)}(t)+(X_{k}(t)-H)\mathds{1}\left\{ Q_{k}(t)>R_{k}(t)-\left(2-(\frac{d}{\tau}-1)A\right)\right\} ,0\right),\label{eq:Virtual1}\\
J_{k}^{(Y)}(t+1) & =\max\left(J_{k}^{(Y)}(t)+(Y_{k}(t)-B)\mathds{1}\left\{ Q_{k}(t)>R_{k}(t)-\left(2-(\frac{d}{\tau}-1)A\right)\right\} ,0\right),\label{eq:Virtual2}\\
J_{k}^{(R)}(t+1) & =\max\left(J_{k}^{(R)}(t)-R_{k}(t)+A,0\right),\label{eq:Virtual3}\\
J_{k}^{(Q)}(t+1) & =\max\left(J_{k}^{(Q)}(t)+R_{k}(t)\mathds{1}\left\{ Q_{k}(t)>R_{k}(t)-\left(2-(\frac{d}{\tau}-1)A\right)\right\} -R_{k}(t)\epsilon_{k},0\right).\label{eq:Virtual4}
\end{flalign}
\noindent\makebox[1\linewidth]{%
\rule{0.84\paperwidth}{0.4pt}%
} 
\end{figure*}

\setcounter{equation}{7}The \emph{Pickands–Balkema–de Haan theorem}
states that for a sufficiently high threshold $\delta$, the distribution
function of the excess value can be approximated by the GPD. In this
regard, considering constraint \eqref{eq:equiv_queue_constraint},
we define the conditional excess queue value of each VUE pair $k\in\mathcal{K}$
at time slot $t$ as $X_{k}(t)|_{Q_{k}(t)>R_{k}(t)-\psi}=Q_{k}(t)-R_{k}(t)+\psi$.
Thus, we can approximate the mean and variance of $X_{k}(t)$ as 
\begin{align}
\mathbb{E}\left[X_{k}(t)|Q_{k}(t)>R_{k}(t)-\psi\right] & \approx\frac{\sigma_{k}}{1-\xi_{k}},\label{eq:Excess_expectation}\\
\text{Var}\left[X_{k}(t)|Q_{k}(t)>R_{k}(t)-\psi\right] & \approx\frac{\sigma_{k}^{2}}{(1-\xi_{k})^{2}(1-2\xi_{k})},\label{eq:Excess_variance}
\end{align}
with a scale parameter $\sigma_{k}$ and a shape parameter $\xi_{k}$.
Note that the smaller the $\sigma_{k}$ and $\xi_{k}$, the smaller
the mean value and variance of the GPD. Hence, we further impose the
thresholds on the scale and the shape parameters, i.e., $\sigma_{k}\leq\sigma_{k}^{th}$
and $\xi_{k}\leq\xi_{k}^{th}$ \cite{Chen-fengMEC}. Subsequently,
applying both parameter thresholds and $\text{Var}\left(X_{k}\right)=\mathbb{E}\left[X_{k}^{2}\right]-\mathbb{E}\left[X_{k}\right]^{2}$
to \eqref{eq:Excess_expectation} and \eqref{eq:Excess_variance},
we consider the constraints for the time-averaged mean and second
moment of the conditional excess queue value, i.e., 
\begin{align}
\hspace{-1em}\bar{X}_{k} & =\lim_{C\rightarrow\infty}\frac{1}{C}{\textstyle \sum\limits _{t=0}^{C-1}}\mathbb{E}\left[X_{k}(t)\rvert Q_{k}(t)>R_{k}(t)-\psi\right]\leq H,\label{eq:GPDMean}\\
\hspace{-1em}\bar{Y}_{k} & =\lim_{C\rightarrow\infty}\frac{1}{C}{\textstyle \sum\limits _{t=0}^{C-1}}\mathbb{E}\left[Y_{k}(t)\rvert Q_{k}(t)>R_{k}(t)-\psi\right]\leq B,\label{eq:GPD2moment}
\end{align}
where $H=\frac{\sigma_{k}^{th}}{1-\xi_{k}^{th}}$, $B=\frac{2(\sigma_{k}^{th})^{2}}{(1-\xi_{k}^{th})(1-2\xi_{k}^{th})}$
and $Y_{k}(t)\coloneqq\left[X_{k}(t)\right]^{2}$. In light of this,
our problem can be formulated as minimizing the total power consumption
of VUEs while ensuring the QoS in terms of reliability constraints
based on AoI. By denoting the RB usage and power allocation vectors
as $\boldsymbol{\eta}(t)=\left[\eta_{k}^{n}(t)\right]_{k\in\mathcal{K}}^{n\in\mathcal{N}}$
and $\boldsymbol{P}(t)=\left[P_{k}^{n}(t)\right]_{k\in\mathcal{K}}^{n\in\mathcal{N}},\forall t$,
respectively, we formulate a network-wide optimization problem which
is written as follows: \begin{subequations}\label{First_optimization_problem}
\begin{align}
\mathbb{P}_{1}:\min_{\boldsymbol{\eta}(t),\boldsymbol{P}(t)} & ~~{\textstyle \sum\limits _{k\in\mathcal{K}}}{\textstyle \sum\limits _{n\in\mathcal{N}}}\bar{P}_{k}^{n}\nonumber \\
\text{subject to} & ~~\eqref{eq:equiv_queue_constraint},\eqref{eq:GPDMean}\text{, and }\eqref{eq:GPD2moment},\nonumber \\
 & ~~\lim_{C\to\infty}\frac{1}{C}{\textstyle \sum\limits _{t=0}^{C-1}}R_{k}(t)>A\text{,}~\forall k\in\mathcal{K},\label{eq:queueStability_const}\\
 & ~~{\textstyle \sum\limits _{n\in\mathcal{N}}}\eta_{k}^{n}(t)P_{k}^{n}(t)\leq P_{\textrm{max}},~\forall k\in\mathcal{K},\label{eq:PowerConst_1}\\
 & ~~0\leq P_{k}^{n}(t)\leq P_{\textrm{max}}\mathds{1}\{\eta_{k}^{n}(t)=1\},\nonumber \\
 & \qquad\qquad\qquad\qquad\forall t,\,k\in\mathcal{K},\,n\in\mathcal{N},\label{eq:PowerConst_2}\\
 & ~~\eta_{k}^{n}(t)\in\{0,1\},~\forall t,\,k\in\mathcal{K},\,n\in\mathcal{N},\label{eq:RBConst}
\end{align}
\end{subequations}where $\bar{P}_{k}^{n}=\lim\limits _{C\to\infty}\frac{1}{C}\sum_{t=0}^{C-1}P_{k}^{n}(t)$
is the time-averaged power consumption of VUE pair $k$ over RB $n$,
and $\mathds{1}\left\{ .\right\} $ is the indicator function. Additionally,
constraint \eqref{eq:queueStability_const} ensures queue stability.
In order to find the optimal resource $\boldsymbol{\eta}(t)$ and
power $\boldsymbol{P}(t)$ allocation vectors that solve problem $\mathbb{P}_{1}$,
we invoke techniques from Lyapunov stochastic optimization \cite{LyapunovNeely2010}.

\section{AoI-Aware Resource Allocation\label{sec:AoI-aware-optimal-control}}

\subsection{Lyapunov Optimization Framework}

We first rewrite \eqref{eq:equiv_queue_constraint} as 
\begin{equation}
\lim_{C\rightarrow\infty}\frac{1}{C}{\textstyle \sum\limits _{t=0}^{C-1}}R_{k}(t)\mathds{1}\left\{ Q_{k}(t)>R_{k}(t)-\psi\right\} \leq\bar{\epsilon}_{k},\label{eq:modified_queueConstraint}
\end{equation}
where $\bar{\epsilon}_{k}=\lim_{C\rightarrow\infty}\frac{1}{C}\sum_{t=0}^{C-1}R_{k}(t)\epsilon_{k}$
is the product of the time-averaged rate and the tolerance value.
Using Lyapunov optimization, the time-averaged constraints \eqref{eq:GPDMean},
\eqref{eq:GPD2moment}, \eqref{eq:queueStability_const}, and \eqref{eq:modified_queueConstraint}
can be satisfied by converting them into virtual queues and maintaining
their stability \cite{LyapunovNeely2010}. In this regard, we introduce
the corresponding virtual queues with the dynamics shown in \eqref{eq:Virtual1}–\eqref{eq:Virtual4}.
Denoting $\boldsymbol{J}(t)=\bigl\{ J_{k}^{(X)}(t),J_{k}^{(Y)}(t),J_{k}^{(Q)}(t),J_{k}^{(R)}(t),Q_{k}(t):k\in\mathcal{K}\bigr\}$
as the combined physical and virtual queue vector, the conditional
Lyapunov drift-plus-penalty for slot $t$ is given by \setcounter{equation}{17}
\begin{equation}
\mathbb{E}\left[\mathscr{L}\left(\boldsymbol{J}(t+1)\right)-\mathscr{L}\left(\boldsymbol{J}(t)\right)+\sum_{k\in\mathcal{K}}\sum_{n\in\mathcal{N}}VP_{k}^{n}(t)\rvert\boldsymbol{J}(t)\right],\label{eq:Lyapunov_drift}
\end{equation}
where $\mathscr{L}\left(\boldsymbol{J}(t)\right)=\frac{\boldsymbol{J'}(t)\boldsymbol{J}(t)}{2}$
is the Lyapunov function. Here, $V\geq0$ is a parameter that controls
the trade-off between power consumption and queue stability. By leveraging
the fact that $\left(\max\left(a-b,0\right)+c\right)^{2}\leq a^{2}+b^{2}+c^{2}-2a(b-c)\text{, }\forall a,b,c\geq0$,
and $\left(\max\left(x,0\right)\right)^{2}\leq x^{2}$ on \eqref{eq:Physical_queue}
and \eqref{eq:Virtual1}–\eqref{eq:Virtual4}, an upper bound on \eqref{eq:Lyapunov_drift}
can be obtained as follows: 
\begin{flalign}
\eqref{eq:Lyapunov_drift} & \leq\mathfrak{C}+\mathbb{E}\Biggl[\sum_{k\in\mathcal{K}}\biggl(\Bigl(J_{k}^{(Q)}(t)-J_{k}^{(X)}(t)-2\left(Q_{k}(t)+\psi\right)^{3}\nonumber \\
 & -\left(2J_{k}^{(Y)}(t)+1\right)\left(Q_{k}(t)+\psi\right)\Bigr)\cdot\mathds{1}\left\{ Q_{k}(t)>R_{k}(t)-\psi\right\} \nonumber \\
 & -\left(J_{k}^{(R)}(t)+A+Q_{k}(t)+J_{k}^{(Q)}(t).\epsilon_{k}\right)\biggr)R_{k}(t)\nonumber \\
 & +\sum_{k\in\mathcal{K}}\sum_{n\in\mathcal{N}}VP_{k}^{n}(t)\rvert\boldsymbol{J}(t)\Biggr].\label{eq:Lyapunov_bound}
\end{flalign}
Here, $\mathfrak{C}$ is a bounded term that does not affect the system
performance.\footnote{Due to the space limitations, the derivations are omitted. The interested
reader may refer to \cite{LyapunovNeely2010} for more details.} Note that the solution to problem $\mathbb{P}_{1}$ can be obtained
by minimizing the upper bound in \eqref{eq:Lyapunov_bound} in each
slot $t$ \cite{LyapunovNeely2010}, i.e., 
\begin{align*}
\mathbb{P}_{2}:\min_{\boldsymbol{\eta}(t),\mathbf{P}(t)} & {\textstyle \sum\limits _{k\in\mathcal{K}}}\Biggl[{\textstyle \sum\limits _{n\in\mathcal{N}}}VP_{k}^{n}(t)-\biggl[J_{k}^{(R)}(t)+A+Q_{k}(t)\\
 & +J_{k}^{(Q)}(t).\epsilon_{k}+\Bigl(-J_{k}^{(Q)}(t)+J_{k}^{(X)}(t)\\
 & +(2J_{k}^{(Y)}(t)+1)(Q_{k}(t)+\psi)+2(Q_{k}(t)+\psi)^{3}\Bigr)\\
 & \times\mathds{1}\left\{ Q_{k}(t)>R_{k}(t)-\psi\right\} \biggr]R_{k}(t)\Biggr]\\
\text{subject to } & \eqref{eq:PowerConst_1}\text{ - }\eqref{eq:RBConst}.
\end{align*}
To solve $\mathbb{P}_{2}$ in each time slot $t$, the RSU needs full
global channel state information (CSI) and queue state information
(QSI). This is clearly impractical for vehicular networks since frequently
exchanging fast-varying local information between the RSU and VUEs
can yield a significant overhead which is not acceptable. To alleviate
the information exchange burden, we propose a two-timescale resource
allocation mechanism which is performed in two stages. Briefly speaking,
RBs are allocated over a long timescale at the RSU whereas each VUE
pair decides its transmit power over a short timescale.

\subsection{Two-Stage Resource Allocation}

\subsubsection{Spectral Clustering and RB Allocation at the RSU}

Before allocating RBs to VUE pairs, we note that if nearby vehicles
transmit over the same RBs, co-channel transmission can lead to severe
interference. In order to avoid the interference from nearby VUEs,
the RSU first clusters VUE pairs into $g>1$ disjoint groups, in which
the nearby VUE pairs are allocated to the same group, and then orthogonally
allocates all RBs to the VUE pairs in each group. Vehicle clustering
is done by means of spectral clustering \cite{Spectral_Clustering}.
We adopt the VUE clustering and RB allocation technique as in \cite{ChenBaseline2018},
denoting $\boldsymbol{v}_{k}\in\mathbb{R}^{2}$ as the Euclidean coordinate
of the midpoint of VUE transmitter-receiver pairs $k$, we use a distance-based
Gaussian similarity matrix $\boldsymbol{S}$ to represent the geographic
proximity information, in which the $(k,k')$-th element is defined
as 
\begin{align*}
s_{kk'}\coloneqq\begin{cases}
e^{-\left\Vert \boldsymbol{v}_{k}-\boldsymbol{v}_{k'}\right\Vert ^{2}/\gamma^{2}}, & \left\Vert \boldsymbol{v}_{k}-\boldsymbol{v}_{k'}\right\Vert \leq\phi,\\
0, & \text{otherwise},
\end{cases}
\end{align*}
where $\phi$ captures the neighborhood size, while $\gamma$ controls
the impact of the neighborhood size. Subsequently, $\boldsymbol{S}$
is used to group VUE pairs using spectral clustering as shown in Algorithm
\ref{alg:Spectral_Clustering}. After forming the groups, the RSU
orthogonally allocates RBs to the VUEs inside the group. Hereafter,
we denote the set $\mathcal{N}_{k}$ as the allocated RB of each VUE
pair $k\in\mathcal{K}$. 
\begin{algorithm}
\begin{algorithmic}[1]

\STATE  Calculate matrix $\boldsymbol{S}$ and the diagonal matrix
$\boldsymbol{D}$ with the diagonal $d_{j}=\sum_{q=1}^{K}s_{jq}$.

\STATE  Let $\boldsymbol{U}=\left[\boldsymbol{u}_{1},\cdots,\boldsymbol{u}_{g}\right]$
in which $\boldsymbol{u}_{g}$ is the eigenvector of the $g$-th smallest
eigenvalue of $\boldsymbol{I}-\boldsymbol{D}^{-\frac{1}{2}}\boldsymbol{S}\boldsymbol{D}^{-\frac{1}{2}}$.

\STATE Numerically, use the $k$-means clustering approach to cluster
$K$ normalized row vectors (which represent $K$ VUE pairs) of matrix
$\boldsymbol{U}$ into $g$ groups.

\end{algorithmic}

\caption{\label{alg:Spectral_Clustering}Spectral Clustering for VUE Grouping}
\end{algorithm}

Moreover, to overcome the signaling overhead issue caused by frequent
information exchange between the RSU and VUE pairs, the RSU clusters
VUE pairs and allocate RBs in a longer time scale, i.e., every $T_{0}\gg1$
time slots, since the geographic locations of the vehicles do not
change significantly during the slot duration $\tau$ (i.e., coherence
time of fading channels). In such, VUE pairs send their geographic
locations to the RSU only once every $T_{0}$ slots instead of every
slot.

\subsubsection{Power Allocation at the VUE}

Since VUE pair $k$ can only use the set $\mathcal{N}_{k}$ of allocated
RBs to send information, we modify the power allocation and RB usage
constraints, i.e., \eqref{eq:PowerConst_1}–\eqref{eq:RBConst}, $\forall k\in\mathcal{K}$,
as 
\begin{align}
\begin{cases}
\sum\limits _{n\in\mathcal{N}_{k}}P_{k}^{n}(t)\leq P_{\textrm{max}},~\forall t,\\
P_{k}^{n}(t)\geq0,~\forall t,\,n\in\mathcal{N}_{k},\\
P_{k}^{n}(t)=0,~\forall t,\,n\notin\mathcal{N}_{k}.
\end{cases}\label{eq:PowerConst_3}
\end{align}
Now the RBs in $\mathcal{N}_{k}$ are reused by distant VUE transmitters
in different groups. We approximately treat the aggregate interference
as a constant term $I$ and rewrite the transmission rate as 
\begin{equation}
\eqref{eq:Rate}\approx\frac{\tau}{Z}{\textstyle \sum\limits _{n\in\mathcal{N}_{k}}}\omega\log_{2}\left(1+\frac{P_{k}^{n}(t)h_{kk}^{n}(t)}{N_{0}\omega+I}\right).\label{eq:rate_2}
\end{equation}
Subsequently, applying \eqref{eq:PowerConst_3} and \eqref{eq:rate_2}
to $\mathbb{P}_{2}$, the VUE transmitter of each VUE pair $k$ locally
allocates its transmit power by solving the following convex optimization
problem $\mathbb{P}_{3}$ in each slot $t$: 
\begin{align*}
\mathbb{P}_{3}:\min_{P_{k}^{n}(t)} & ~~{\textstyle \sum\limits _{n\in\mathcal{N}_{k}}}VP_{k}^{n}(t)-\Im_{k}(t)\log_{2}\left(1+\frac{P_{k}^{n}(t)h_{kk}^{n}(t)}{N_{0}\omega+I}\right)\\
\text{subject to} & ~~\eqref{eq:PowerConst_3},
\end{align*}
where $\Im_{k}(t)=\frac{\tau\omega}{Z}\biggl[J_{k}^{(R)}(t)+A+Q_{k}(t)+J_{k}^{(Q)}(t)\epsilon_{k}+\Bigl(-J_{k}^{(Q)}(t)+J_{k}^{(X)}(t)+(2J_{k}^{(Y)}(t)+1)(Q_{k}(t)+\psi)+2(Q_{k}(t)+\psi)^{3}\Bigr)\times\mathds{1}\left\{ Q_{k}(t)>R_{k}(t)-\psi\right\} \biggr]$.
Based on the Karush-Kuhn-Tucker (KKT) conditions, the optimal VUE
transmit power $P_{k}^{n*}(t),\forall n\in\mathcal{N}_{k},$ of $\mathbb{P}_{3}$
satisfies, 
\[
\frac{\Im_{k}(t)h_{kk}^{n}(t)}{(N_{0}\omega+I+P_{k}^{n*}(t)h_{kk}^{n}(t))\ln2}=V+\zeta,
\]
if $\frac{\Im_{k}(t)h_{kk}^{n}(t)}{(N_{0}\omega+I)\ln2}>V+\zeta.$
Otherwise, $P_{k}^{n*}(t)=0.$ Moreover, the Lagrange multiplier $\zeta$
is $0$ if $\sum_{n\in\mathcal{N}_{k}}P_{k}^{n*}(t)<P_{\textrm{max}}$,
and we have $\sum_{n\in\mathcal{N}_{k}}P_{k}^{n}(t)=P_{\textrm{max}}$
when $\zeta>0$ . Note that, given a small value of $V$, the derived
power $P_{k}^{n*}(t)$ provides a sub-optimal solution to problem
$\mathbb{P}_{1}$, whose optimal solution is asymptotically obtained
by increasing $V$. After sending its status update, VUE pair $k$
updates the physical and virtual queues as per \eqref{eq:Physical_queue}
and \eqref{eq:Virtual1}–\eqref{eq:Virtual4}.

\section{Simulation Results and Analysis\label{sec:Numerical-results}}

\begin{table}[t]
\caption{Simulation parameters \cite{ChenBaseline2018}.\label{tab:sim_par}}
\centering{}%
\begin{tabular}{|c|c||c|c|}
\hline 
\textbf{Parameter}  & \textbf{Value}  & \textbf{Parameter}  & \textbf{Value}\tabularnewline
\hline 
\hline 
$N$  & $20$  & $H$  & $0.8334$\tabularnewline
\hline 
$\omega$  & $180$ KHz  & $B$  & $0.7576$\tabularnewline
\hline 
$\tau$  & $3$ ms  & $g$  & $10$\tabularnewline
\hline 
$P_{\textrm{max}}$  & $23$ dBm  & $\gamma$  & $30$ m\tabularnewline
\hline 
$Z$  & $500$ Byte  & $\phi$  & $150$ m\tabularnewline
\hline 
$N_{0}$  & $-174$ dBm/Hz  & $T_{0}$  & $100$\tabularnewline
\hline 
Arrival rate  & $0.5$ Mbps  & $\alpha$  & $1.61$\tabularnewline
\hline 
$d$  & $60$ ms  & $\mathscr{D}$  & $15$ m\tabularnewline
\hline 
$\epsilon_{k}$  & $0.001$  & $l_{0}$  & $-68.5$ dB\tabularnewline
\hline 
$\psi$  & $-3.25$  & $l_{0}^{'}$  & $-54.5$ dB\tabularnewline
\hline 
$V$  & $0$  &  & \tabularnewline
\hline 
\end{tabular}
\end{table}

For our simulations, we use a $250\times250\text{ m}^{2}$ area Manhattan
mobility model as in \cite{Ikram2017,ChenBaseline2018}. The average
vehicle speed is $60\text{ km/h}$, and the distance between the transmitter
and receiver of each VUE pair is $15\text{ m}$. Unless stated otherwise,
the remaining parameters are listed in Table \ref{tab:sim_par}. The
performance of our proposed solution is compared to that of \cite{ChenBaseline2018},
where a power minimization is considered subject to a reliability
measure in terms of maximal queue length among all VUEs.

\begin{figure}
\begin{centering}
\centering\includegraphics[scale=0.65]{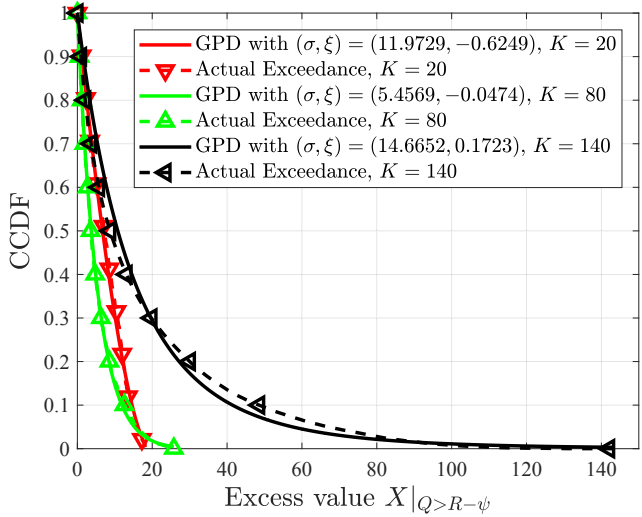} 
\par\end{centering}
\caption{CCDF of the exceedance value fitted to GPD for various densities of
VUEs ($K$).\label{fig:fitting}}
\end{figure}

In Fig.~\ref{fig:fitting}, we verify the accuracy of using EVT to
characterize the distribution of the excess value $X_{k}(t)|_{Q_{k}(t)>R_{k}(t)-\psi}=Q_{k}(t)-R_{k}(t)+\psi$.
In particular, Fig.~\ref{fig:fitting} shows the complementary cumulative
distribution function (CCDF) of the excess value for various densities
of VUEs, along with the GPD distributions with the specified parameters.
A near-perfect fitting can be noted, which verifies the accuracy of
using EVT to characterize the distribution of the excess value.

\begin{figure}
\begin{centering}
\centering\includegraphics[scale=0.64]{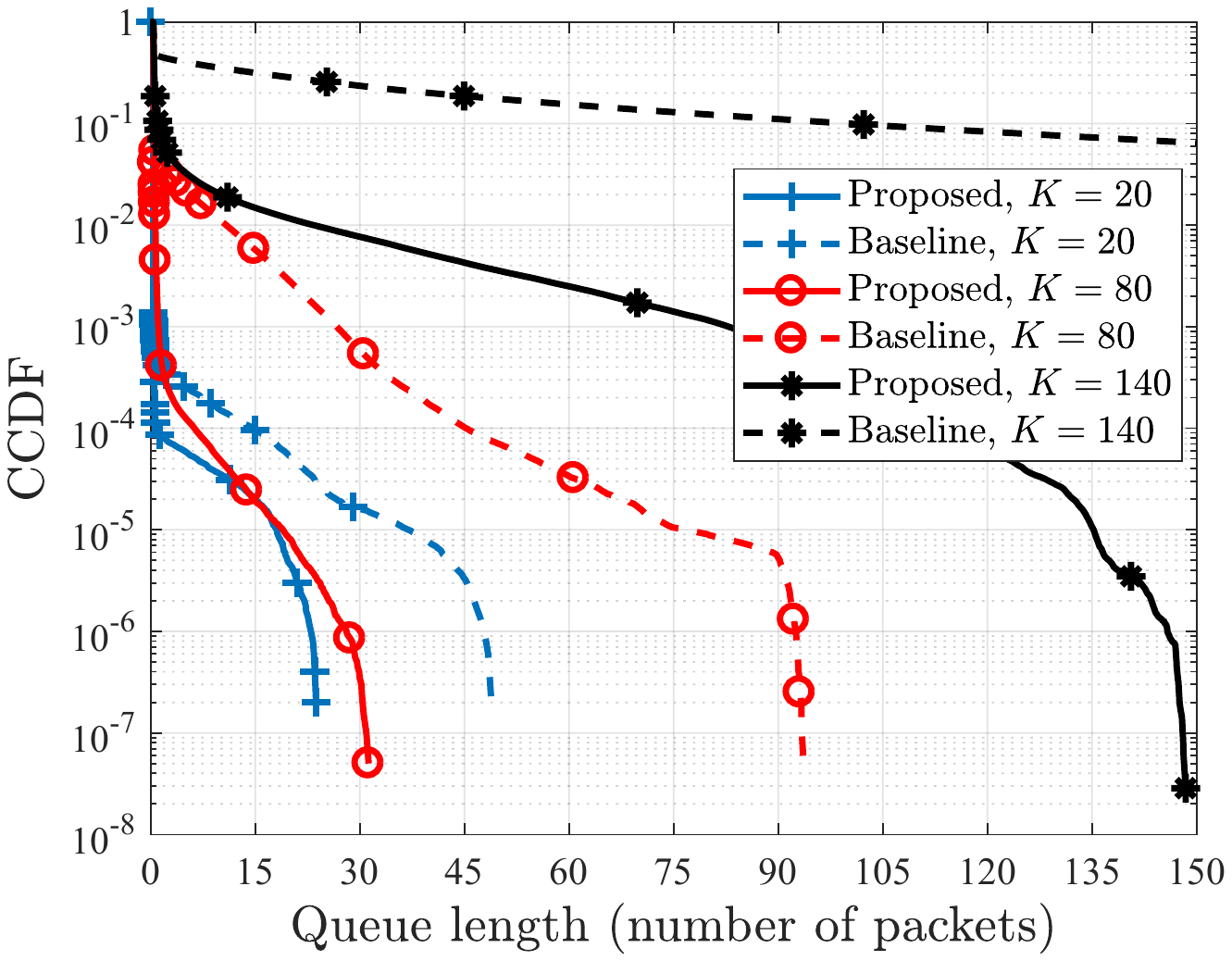} 
\par\end{centering}
\caption{CCDF of the queue length for various densities of VUEs $(K)$.\label{fig:CCDF_Queue}}
\end{figure}

\begin{figure}
\begin{centering}
\centering\includegraphics[scale=0.64]{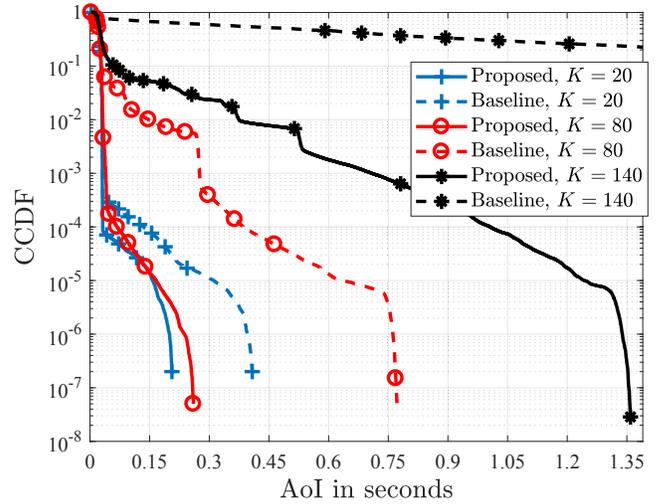} 
\par\end{centering}
\caption{CCDF of the AoI for various densities of VUEs ($K$).\label{fig:CCDF_AoI}}
\end{figure}

\begin{figure}
\begin{centering}
\centering\includegraphics[scale=0.64]{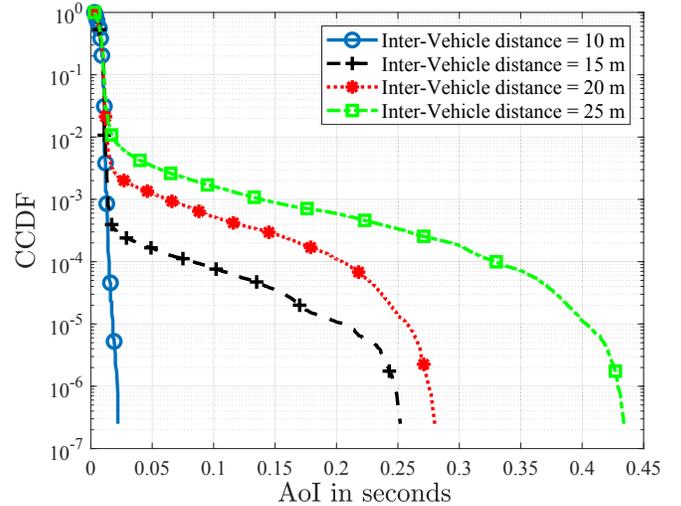} 
\par\end{centering}
\caption{CCDF of the AoI for various inter-vehicle distance with $K=80$.\label{fig:CCDF_distance}}
\end{figure}

In Fig.~\ref{fig:CCDF_Queue}, we show the CCDF of the queue length
of all VUEs, in which our approach achieves a better distribution
in terms of the tail, compared to the baseline with a reduction in
the worst queue length by more than two-fold for all cases. The AoI
distribution, which is the main reliability measure in our model is
shown in Fig.~\ref{fig:CCDF_AoI} in terms of its CCDF for various
densities of VUEs. Again, the AoI distribution in our approach outperforms
that of the baseline with a much shorter tail (higher reliability).
By inspecting both Fig.~\ref{fig:CCDF_Queue} and Fig.~\ref{fig:CCDF_AoI},
we can observe that both queue length and AoI distributions follow
the same trend, which verify the proposed mapping between the probabilistic
AoI and the queue length (Lemma \ref{lem:Assuming-that-packet }).
It is worth mentioning that, these sudden changes in the distributions
are due to the adopted path loss model, ranging from LOS, WLOS and
NLOS. To further validate this, we have varied the distance between
the transmitter and receiver of each VUE pair from $10$ m to $25$
m. Fig.~\ref{fig:CCDF_distance} shows that the tail of AoI distribution
decreases as the inter-vehicle distance decreases until it vanishes
at a $10$ m distance. 

\begin{figure}
\begin{centering}
\centering\includegraphics[scale=0.65]{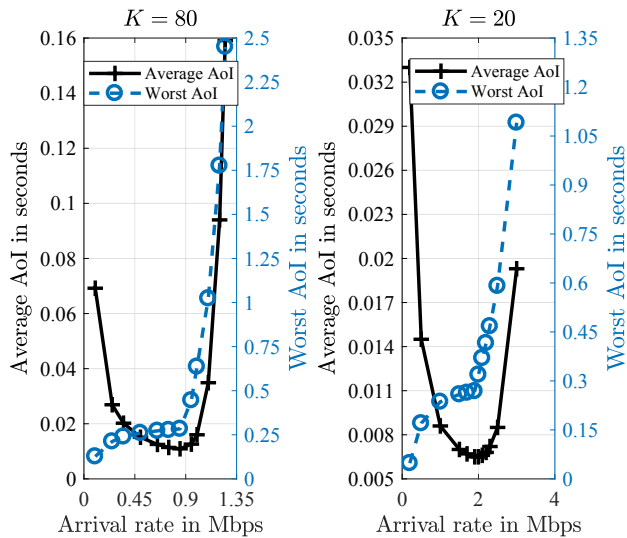} 
\par\end{centering}
\caption{Arrival rate versus AoI trade-off, with $K=80$ and $K=20$ VUEs.\label{fig:tradeoff}}
\end{figure}

Finally, in Fig.~\ref{fig:tradeoff} we study the impact of the arrival
rate of the status updates for two different VUE densities, $K=80$
and $K=20$, for both average and worst AoI. When the arrival rate
is small, the network incurs higher average AoI due to the lack of
status updates sent by VUEs. Hence, increasing the arrival rate will
reduce that lack of status updates, leading to a better average AoI,
up to a certain point, after which the average AoI starts increasing
again due to interference. Note that, at low VUE density ($K=20$)
the network can withstand higher arrival rates. Fig.~\ref{fig:tradeoff}
also highlights that the increase of the worst AoI with the arrival
rate shows a slight increase up to the aforementioned optimal arrival
rate for the average AoI, and exhibits a rapid increase afterwards.
Moreover, we note that the worst AoI is about 10 fold higher than
the average AoI. Although the AoI achieves a small average, the worst
AoI is heavy-tailed. In this situation, relying on the average AoI
is inadequate for ensuring URLLC. This discrepancy between these two
metrics demonstrates that the tail characterization is instrumental
in designing and optimizing URLLC-enabled V2V networks. 

\section{Conclusion\label{sec:Conclusion}}

In this paper, we have studied the problem of ultra-reliable and low-latency
vehicular communication. For this purpose, we have defined a new reliability
measure, in terms of probabilistic AoI. Second, we have established
a novel relationship between the AoI and queue length probability
distributions. Then, we have shown that characterizing the queue length
tail distribution can be effectively done using EVT. Subsequently,
we formulated the problem as a power minimization problem subject
to the probabilistic AoI and solved it using Lyapunov optimization.
Simulation results show that the proposed approach yields significant
improvements in terms of AoI and queue length, when compared to a
baseline that only considers the maximum queue among vehicles. In
our future work, we will extend the current framework to other use
cases such as autonomous aerial vehicles (UAVs) \cite{Mozaffari2016,7875131}
and industrial control \cite{Industrial}. Another interesting extension
pertains to studying the impact of short packets and finite blocklength
on our framework and its performance \cite{Polyanskiy2010}.

\section*{Acknowledgment}

This work was supported in part by the Academy of Finland project
CARMA, and 6Genesis Flagship (grant no. 318927), in part by the INFOTECH
project NOOR, in part by the U.S. National Science Foundation under
Grants CNS-1513697 and CNS-1739642, and in part by the Kvantum Institute
strategic project SAFARI.

\bibliographystyle{IEEEtran}
\bibliography{CameraReady}

\begin{thebibliography}{10}
\providecommand{\url}[1]{#1}
\csname url@samestyle\endcsname
\providecommand{\newblock}{\relax}
\providecommand{\bibinfo}[2]{#2}
\providecommand{\BIBentrySTDinterwordspacing}{\spaceskip=0pt\relax}
\providecommand{\BIBentryALTinterwordstretchfactor}{4}
\providecommand{\BIBentryALTinterwordspacing}{\spaceskip=\fontdimen2\font plus
\BIBentryALTinterwordstretchfactor\fontdimen3\font minus
  \fontdimen4\font\relax}
\providecommand{\BIBforeignlanguage}[2]{{%
\expandafter\ifx\csname l@#1\endcsname\relax
\typeout{** WARNING: IEEEtran.bst: No hyphenation pattern has been}%
\typeout{** loaded for the language `#1'. Using the pattern for}%
\typeout{** the default language instead.}%
\else
\language=\csname l@#1\endcsname
\fi
#2}}
\providecommand{\BIBdecl}{\relax}
\BIBdecl

\bibitem{Araniti2013}
G.~Araniti, C.~Campolo, M.~Condoluci, A.~Iera, and A.~Molinaro, ``{LTE} for
  vehicular networking: A survey,'' \emph{IEEE Commun. Mag.}, vol.~51, no.~5,
  pp. 148--157, May 2013.

\bibitem{Zeng2018}
T.~Zeng, O.~Semiari, W.~Saad, and M.~Bennis, ``Joint communication and control
  for wireless autonomous vehicular platoon systems,'' \emph{CoRR}, vol.
  arXiv/1804.05290, 2018.

\bibitem{8255748}
S.~A.~A. Shah, E.~Ahmed, M.~Imran, and S.~Zeadally, ``5g for vehicular
  communications,'' \emph{IEEE Communications Magazine}, vol.~56, no.~1, pp.
  111--117, Jan. 2018.

\bibitem{etsi2011intelligent}
{ETSI EN Std 302 637-2}, ``Intelligent transport systems; vehicular
  communications; basic set of applications; part 2: Specification of
  cooperative awareness basic service,'' Aug. 2013.

\bibitem{Mehdi_riskTail}
M.~Bennis, M.~Debbah, and H.~V. Poor, ``Ultra-reliable and low-latency wireless
  communication: Tail, risk and scale,'' \emph{Proc. IEEE}, 2018, to be
  published.

\bibitem{EVT}
S.~Coles, \emph{An introduction to statistical modeling of extreme
  values}.\hskip 1em plus 0.5em minus 0.4em\relax Springer, 2001.

\bibitem{Ashraf2016}
M.~I. Ashraf, M.~Bennis, C.~Perfecto, and W.~Saad, ``Dynamic proximity-aware
  resource allocation in vehicle-to-vehicle ({V2V}) communications,'' in
  \emph{Proc. IEEE Global Commun. Conf. Workshops}, Dec. 2016, pp. 1--6.

\bibitem{Ikram2017}
M.~I. Ashraf, C.-F. Liu, M.~Bennis, and W.~Saad, ``Towards low-latency and
  ultra-reliable vehicle-to-vehicle communication,'' in \emph{Proc. Eur. Conf.
  Netw. Commun.}, Jun. 2017, pp. 1--5.

\bibitem{Liu2017}
T.~Liu, Y.~Zhu, R.~Jiang, and Q.~Zhao, ``Distributed social welfare
  maximization in urban vehicular participatory sensing systems,'' \emph{IEEE
  Trans. Mobile Comput.}, vol.~17, no.~6, pp. 1314--1325, Jun. 2018.

\bibitem{Mei2018}
J.~Mei, K.~Zheng, L.~Zhao, Y.~Teng, and X.~Wang, ``A latency and reliability
  guaranteed resource allocation scheme for {LTE} {V2V} communication
  systems,'' \emph{IEEE Trans. Wireless Commun.}, vol.~17, no.~6, pp.
  3850--3860, Jun. 2018.

\bibitem{Cristina}
C.~Perfecto, J.~D. Ser, and M.~Bennis, ``Millimeter-wave v2v communications:
  Distributed association and beam alignment,'' \emph{IEEE Journal on Selected
  Areas in Communications}, vol.~35, no.~9, pp. 2148--2162, Sep. 2017.

\bibitem{Mouradian2016}
A.~Mouradian, ``Extreme value theory for the study of probabilistic worst case
  delays in wireless networks,'' \emph{Ad Hoc Networks}, vol.~48, pp. 1--15,
  Sep. 2016.

\bibitem{ChenBaseline2018}
C.-F. Liu and M.~Bennis, ``Ultra-reliable and low-latency vehicular
  transmission: An extreme value theory approach,'' \emph{IEEE Commun. Lett.},
  vol.~22, no.~6, pp. 1292--1295, Jun. 2018.

\bibitem{champatiAoI}
J.~Champati, H.~Al-Zubaidy, and J.~Gross, ``Statistical guarantee optimization
  for age of information for the {D/G/1} queue,'' in \emph{Proc. IEEE Conf.
  Comput. Commun. Workshops}, Apr. 2018, pp. 130--135.

\bibitem{AoI_First}
S.~Kaul, M.~Gruteser, V.~Rai, and J.~Kenney, ``Minimizing age of information in
  vehicular networks,'' in \emph{Proc. 8th Annual IEEE Communications Society
  Conference on Sensor, Mesh and Ad Hoc Communications and Networks}, Jun.
  2011, pp. 350--358.

\bibitem{AoIBaiocchi2017}
A.~Baiocchi and I.~Turcanu, ``A model for the optimization of beacon message
  age-of-information in a {VANET},'' in \emph{Proc. IEEE 29th Int. Teletraffic
  Congress}, Sep. 2017, pp. 108 -- 116.

\bibitem{Zhou2018}
B.~Zhou and W.~Saad, ``Joint status sampling and updating for minimizing age of
  information in the internet of things,'' \emph{CoRR}, vol. arXiv/1807.04356,
  2018.

\bibitem{AoIAbd-Elmagid2018}
M.~A. Abd-Elmagid and H.~S. Dhillon, ``Average age-of-information minimization
  in {UAV}-assisted {IoT} networks,'' \emph{CoRR}, 2018.

\bibitem{Manhattan}
F.~Bai, N.~Sadagopan, and A.~Helmy, ``{IMPORTANT}: A framework to
  systematically analyze the impact of mobility on performance of routing
  protocols for adhoc networks,'' in \emph{Proc. IEEE Conf. Comput. Commun.},
  vol.~2, Mar. 2003, pp. 825--835.

\bibitem{Path_loss_model}
M.~Abdulla and H.~Wymeersch, ``Fine-grained vs. average reliability for {V2V}
  communications around intersections,'' in \emph{Proc. IEEE Global Commun.
  Conf. Workshops}, Dec. 2017, pp. 1--7.

\bibitem{Chen-fengMEC}
C.-F. Liu, M.~Bennis, and H.~V. Poor, ``Latency and reliability-aware task
  offloading and resource allocation for mobile edge computing,'' in
  \emph{Proc. IEEE Global Commun. Conf. Workshops}, Dec. 2017, pp. 1--7.

\bibitem{LyapunovNeely2010}
M.~J. Neely, \emph{Stochastic Network Optimization with Application to
  Communication and Queueing Systems}.\hskip 1em plus 0.5em minus 0.4em\relax
  Morgan and Claypool Publishers, Jun. 2010.

\bibitem{Spectral_Clustering}
U.~von Luxburg, ``A tutorial on spectral clustering,'' \emph{Statistics and
  Computing}, vol.~17, no.~4, pp. 395--416, Dec. 2007.

\bibitem{Mozaffari2016}
M.~Mozaffari, W.~Saad, M.~Bennis, and M.~Debbah, ``Unmanned aerial vehicle with
  underlaid device-to-device communications: Performance and tradeoffs,''
  \emph{IEEE Trans. Wireless Commun.}, vol.~15, no.~6, pp. 3949--3963, Jun.
  2016.

\bibitem{7875131}
M.~Chen, M.~Mozaffari, W.~Saad, C.~Yin, M.~Debbah, and C.~S. Hong, ``Caching in
  the sky: Proactive deployment of cache-enabled unmanned aerial vehicles for
  optimized quality-of-experience,'' \emph{IEEE Journal on Selected Areas in
  Communications (JSAC), Special Issue on Human-In-The-Loop Mobile Networks},
  vol.~35, no.~5, pp. 1046--1061, May 2017.

\bibitem{Industrial}
O.~N.~C. Yilmaz, Y.~.~E. Wang, N.~A. Johansson, N.~Brahmi, S.~A. Ashraf, and
  J.~Sachs, ``Analysis of ultra-reliable and low-latency 5{G} communication for
  a factory automation use case,'' in \emph{2015 IEEE International Conference
  on Communication Workshop (ICCW)}, Jun. 2015, pp. 1190--1195.

\bibitem{Polyanskiy2010}
Y.~Polyanskiy, H.~V. Poor, and S.~Verdu, ``Channel coding rate in the finite
  blocklength regime,'' \emph{IEEE Trans.~Inf.~Theory}, vol.~56, no.~5, pp.
  2307--2359, May 2010.

\end{thebibliography}

\end{document}